# Solid State Thermionic Power Generators: an analytical analysis in the nonlinear regime

M. Zebarjadi,[1,2,a]

[1] *Department of Electrical and Computer Engineering, University of Virginia, Charlottesville, VA, 22904, USA*
[2] *Department of Materials and Science Engineering, University of Virginia, Charlottesville, VA, 22904, USA*

Solid-state thermionic power generators are an alternative to thermoelectric modules. In this manuscript, we develop an analytical model to investigate the performance of these generators in the non-linear regime. We identify dimensionless parameters determining their performance and provide measures to estimate acceptable range of thermal and electrical resistances of thermionic generators. We find the relation between the optimum load resistance and the internal resistance and suggest guide lines for the design of thermionic power generators. Finally, we show that in the nonlinear regime, thermionic power generators can have efficiency values higher than the state of the art thermoelectric modules.

**I. INTRODUCTION:**

Solid-state thermionic power generators can be viewed as a bridge between vacuum-state thermionic convertors and thermoelectric power generators. Like many other heat engines, these devices can work either as power generators where they convert input heat to electricity, or coolers where applied electricity is used to pump heat. Solid-state thermionic coolers and power generators were first proposed by Mahan[1,2] and Shakouri.[3,4] The working principles of solid-state thermionic convertors is very similar to vacuum-state thermionic convertors. In vacuum-state thermionic conversion, a cathode and an anode are separated by vacuum. Electrons in cathode are heated to high energies. When they gain enough velocity in the out of plane direction to overcome the energy barrier (cathode's work function), they leave the cathode and enter the vacuum. Some of these electrons are eventually collected at the anode side and flow through the outer circuit. Vacuum thermionic diodes were invented in 1904 and were used in radio and telephone communications. However, it was only in 1956 that two distinct practical versions of thermionic power generators were made by Wilson[5] and Hatsopoulos[6]. Efficiencies as high as 16% and large power densities reaching $11 W\ cm^{-2}$ were reported in early demonstrations.[7]

The main drawback of vacuum-state thermionic convertors is that they can only operate at high temperatures (>1000K). In solid-state thermionic convertors, the vacuum is replaced by a solid, usually a semiconductor. Solid-state thermionic convertors have several advantages to vacuum-state thermionic convertors. First, the effective barrier height is the difference between the cathode's work function and the semiconductor's electron affinity. Therefore, it is easy to lower the barrier height and extend the operation temperature to low temperatures. Second, the absence of vacuum makes the fabrication process and the access to the cathode and the anode for the purpose of heating and cooling, easier. Finally, vacuum thermionic convertors suffer from the space charge effect. This effect is negligible in solid-state thermionic convertors due to the extremely small size of the barrier region and the fact that the chemical potential of the semiconductors could be tuned by means of doping to minimize the band bending (which is the equivalent of space charge effect in vacuum-state devices). The main disadvantage of the solid-state thermionic convertors is their heat leakage by conduction through the semiconductor which is more severe compared to thermal radiation in vacuum-state thermionic convertors. This was noted from the very beginning by Mahan.[2] Mahan then proposed that only small temperature differences should be imposed to each barrier and multi-barrier structures should be used. Shakouri proposed that tall barriers in the electron path could be used to filter high energy electrons and increase the average energy per carrier to increase the efficiency.[3]

In this work, we develop an analytical model to investigate the operation limits of solid-state thermionic power generators. We focus on tall barriers and large temperature differences. The linear regime of small temperature differences has been investigated before by Vining and Mahan. They showed that in the linear limit, thermoelectric convertors outperform thermionic convertors.[8]

---

[a] Author to whom correspondence should be addressed. Electronic mail: m.zebarjadi@virginia.edu.

**II. Model:**

Consider a single barrier structure. Cathode and anode are separated by a semiconductor. The semiconductor thickness is smaller than the electron mean-free-path, ensuring ballistic transport. We define an internal resistance for the structure, R. Ideally this internal resistance should be zero. In practice it is never zero and even for a purely ballistic channel, has the contribution of the cathode and anode resistances, the interfacial resistances, and even outer circuit connecting wire resistances. Hillary Moss was the first to investigate the importance of the internal resistance on the performance of the thermionic generators.[9] In the context of vacuum-state thermionic generators, she concluded that even a small internal resistance (fraction of an ohm) drops the output power.

We also define a thermal resistance, $R_t$, which includes the thermal resistance of the barrier layer and the interfacial resistance between the barrier and the cathode and the anode layers. Ideally, this resistance should be infinitely large. Since the thickness of the barrier region is small, its thermal resistance is also small. Therefore, a good design requires large interfacial thermal resistances. Assuming the barrier is tall, the electron population above the barrier could be approximated by the Maxwellian distribution and the current could be described by Richardson-Dushman current. In a typical vacuum thermionic convertor, cathode's work function is larger than the anode's work function. There are two distinct operating regions: 1- when the cathode's work function, $\phi_c$, is larger than the anode's work function, $\phi_a$, plus the total voltage ($V = IR + IR_l$, $R_l$ is the load resistance) and 2- the opposite case when $\phi_c < \phi_a + V$

It can be easily shown that in the first case, the current is independent of $\phi_a$, and in the second case it is independent of $\phi_c$.

In case 1, the reverse current from anode to cathode is small and current is almost constant with respect to voltage, while in case 2, current drops exponentially with respect to voltage. We write our equations for case 1 and we assume that the optimum power with respect to current/voltage happens in this regime. The electrical and the thermal current, the power, and the efficiency for ($\phi_c > \phi_a + V$) can be written as:[1,9]

$$J = A_R T_c^2 \exp\left(-\frac{e\phi_c}{k_B T_c}\right) - A_R T_a^2 \exp\left(-\frac{e(\phi_c-V)}{k_B T_a}\right) \qquad (1)$$

$$J_{Qc} = A_R T_c^2 \exp\left(-\frac{e\phi_c}{k_B T_c}\right)(\phi_c + 2k_B T_c/e) - A_R T_a^2 \exp\left(-\frac{e(\phi_c-V)}{k_B T_a}\right)(\phi_c + 2k_B T_a/e) + \frac{T_c-T_a}{R_t} \qquad (2)$$

$$p = \frac{P}{A} = IR_l J = (V - IR)J \qquad (3)$$

$$\eta = \frac{p}{J_{Qc}} \qquad (4)$$

$J$ is the electrical current; $J_{Qc}$ is the thermal current at the cathode/barrier junction; $p$ is the power density dissipated in the load, and $\eta$ is the efficiency of thermal to electrical energy conversion. $T_c$ is the cathode temperature (hot), $T_a$ is the anode temperature (cold), $k_B$ is the Boltzmann constant, $e$ is the electric charge, and $A_R$ is the Richardson constant containing effective masses of the cathode and the anode and the average transmission function. To have zero current at zero voltage and zero temperature gradient, $A_R$ should be the same for both cathode and anode current.

The first term in Eq. 1 is the current from cathode to anode and the second term is that from anode to cathode. Similarly, in Eq. 2 the first term is the thermal current from cathode to anode and the second term is the back flow from anode to cathode. The third term in Eq.2 is the lattice conduction term or the heat leak via barrier's lattice and from cathode to anode. Here we neglect joule heating inside the barrier as we assumed ballistic transport. Note that there is joule heating outside of the barrier, in the cathode and the anode region but that is not of interest and as long as the electron-phonon mean free path is larger than the barrier thickness, joule heating within the barrier can be ignored.

To make the analysis simple, we define dimensionless parameters as listed below:

$$\Phi_c = \frac{e\phi_c}{k_B T_c}; \Phi_a = \frac{e\phi_a}{k_B T_a}; \quad V_a = \frac{eV}{k_B T_a}; \theta = \frac{T_c}{T_a}; J_D = \frac{J}{A_R T_a^2}; J_{QD} = \frac{eJ_Q}{k_B T_a A_R T_a^2}; P_D = \frac{ep}{k_B T_a A_R T_a^2} \quad (5)$$

We use subscript D to refer to dimensionless current and power. Now we can rewrite Eqs. 1 to 4, in a dimensionless format:

$$J_D = \theta^2 \exp(-\Phi_c) - \exp(-\Phi_c \theta + V_a) = J_c - J_0 \exp(V_a) \quad (6)$$

$$J_a = J_0 \exp(V_a) \quad (7)$$

$$J_c = \theta^2 \exp(-\Phi_c) \quad (8)$$

$$J_0 = \exp(-\Phi_c \theta) \quad (9)$$

$$V_a = \ln\left(\frac{J_c - J_D}{J_0}\right) \quad (10)$$

$$J_{QD} = J_c(\Phi_c \theta + 2\theta) - J_0 \exp(V_a)(\Phi_c \theta + 2) + \frac{\theta - 1}{R_{tD}} \quad (11)$$

$$R_{tD} = R_t k_B A_R T_a^2 / e \quad (12)$$

$$P_D = (V_a - J_D R_D) J_D \quad (13)$$

$$R_D = \frac{A_R T_a A R e}{k_B} \quad (14)$$

$R_{tD}$ is the dimensionless internal thermal resistance and the $R_D$ is the dimensionless internal electrical resistance.

Similar equations could be written for the second case where $\phi_c < \phi_a + V$.

These equations could be solved numerically. However, it is the purpose of this work to find analytical solutions. Before we move on to analytical solutions, let us first look at numerical results to develop an understanding of the performance of the device. Figure 1 shows the numerical results for a given set of parameters. Values of the parameters used, are shown in the caption of Fig.1.

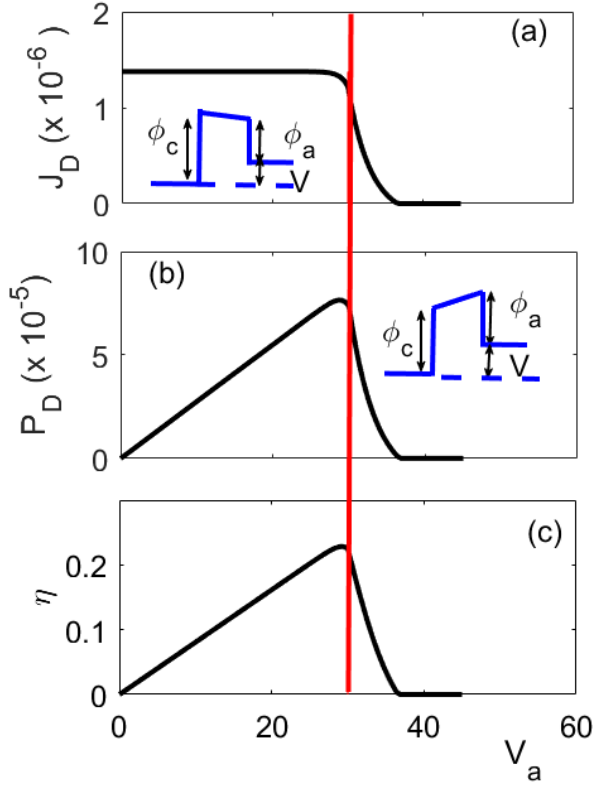

FIG. 1. Dimensionless current, power and efficiency versus voltage for $\theta = 3; R_D = R_{tD} = 10,000; \Phi_c = 15; \Phi_a = 15$. The red line ($V_a = \Phi_c\theta - \Phi_a$) is separating the two operating regions discussed in the text. A schematic of the band diagram in case 1 is shown as an inset of plot (a) and a schematic of the band diagram in case 2 is shown as an inset of plot (b).

**III. Results:**

Now let us discuss the optimum performance of the device analytically. We note that the optimum power and the optimum efficiency do not happen at exactly the same current/voltage but at very close ones. We first optimize the power with respect to current. Taking the derivative of Eq. 13 with respect to current and setting it to zero, we find the optimal condition to be:

$$\left(\frac{1}{J_c-J_D} + 2R_D\right)J_D = V_a = J_D(R_D + R_{lD}) \tag{15}$$

Therefore, when power is maximum, the following relation holds between the load resistance and the internal resistance.

$$\frac{1}{J_a} + R_D = R_{lD} \tag{16}$$

To find the optimum current we use Eq. 15 and Eq. 10 and we find:

$$J_c + (2J_cR_D - 1)J_a - 2R_DJ_a^2 = J_a \ln\left(\frac{J_a}{J_0}\right) \tag{17}$$

This equation does not have an analytical solution. However, the second order term in $J_a$ is small and can be neglected. We then find at the optimum power:

$$J_a \approx \frac{J_c}{LW(x_0)} \tag{18}$$

$$x_0 = \theta^2 \exp\left((\theta - 1)\Phi_c + 1 - 2J_cR_D\right) \tag{19}$$

Here LW is the Lambert W function (See appendix A ).

Finally, the optimum current, the corresponding load resistance, the optimum power and the corresponding efficiency can be written as

$$J_D = J_c y \tag{20}$$

$$y = 1 - \frac{1}{LW(x_0)} \qquad (21)$$

$$\frac{LW(x_0)}{J_c} + R_D = R_l \qquad (22)$$

$$P_{D_{opt}} = (J_C R_D + LW(x_0)) J_c y^2 \qquad (23)$$

$$\eta = \frac{(J_c R_D + LW(x_0)) y^2}{(\Phi_c \theta + 2) y + (2 + \frac{1}{J_c R_{tD}})(\theta - 1)} \qquad (24)$$

Two immediate conclusions can be drawn regarding the acceptable range of the internal resistances. For electrical resistance, noting that $(\theta - 1)\Phi_c > 0$, we find that if $2J_c R_D \ll 1$, then the effect of internal resistance is very small (Eq. 19), and does not substantially drop the power. The criteria can be written as $R \ll 0.5 \frac{k_B T_a}{e J_R A}$, where $J_R$ is the Richardson current (in units of Amp) from cathode to anode. If we use the theoretical Richardson constant of $120\ A\ cm^{-2} K^{-2}$; barrier height, $\Phi_c$, of 5; cathode and anode area of about $1\ cm^2$; $T_a$ of 300K, and $T_c = 400K$, we find $R \ll 1 \times 10^{-7} \Omega$ ($R_D \sim 44$), which means any internal resistance larger than this value will lower the output power from that of the ideal thermionic convertor. The resistance of the barrier layer is much smaller since the thickness is small (less than 100nm). The resistance of the cathode and anode could also be kept small. The resistance of a typical metal of thickness 1mm and area of $1\ cm^2$, is $R \sim 10^{-8} \Omega m\ 10^{-3} m / 10^{-4} m^2 \sim 10^{-7} \Omega$, which means the internal resistance of the solid-state thermionic generators could be kept really small to have a minimized effect on the output power.

Similarly, for the thermal resistance, we can conclude that if $J_c R_{tD} \gg 1$, then the lattice thermal conduction is minimal and would not substantially drop the efficiency. This condition is equivalent to $R_t \gg \frac{e}{k_B J_R}$. Using the same parameters as before, we find $R_t \gg 9 \times 10^{-6}\ m^2 K\ W^{-1}$ or $G_t \ll 0.1\ MW m^{-2} K^{-1}$ which is extremely small. If we take a typical material with thermal conductivity of $1\ W m^{-1} K^{-1}$ and thickness of $100nm$, then the equivalent $G_t = \frac{\kappa}{L} = 10\ MW m^{-2} K^{-1}$. Of course if we increase the thickness, we can lower the conductance to an arbitrarily low value. However, we note that the criteria for thermionic emission is that the thickness should be smaller than the electron mean-free-path. Therefore, it is not possible to lower the conductance without breaking the thermionic conditions. We believe the only possible solution is to use weakly-bonded materials. One example is layered van der Waals heterostructures. In a recent work,[10] we have shown that for layered materials one can achieve very low conductance values. For only 5 layers of black phosphorene (2.5nm length) sandwiched between gold contacts, we have calculated a thermal conductance of $4\ MW m^{-2} K^{-1}$ for an ideal (defect-free) case. The value could be substantially smaller in practice due to imperfections. One can further decrease it, by purposely introducing defects at the interfaces. As long as the defects do not block electrical transport and do not break the ballistic transport condition inside the barrier region, such approach is an acceptable way of increasing the interfacial thermal resistance and therefore the efficiency. Increasing the thermal resistance while maintaining a low electrical resistance remains the biggest challenge for solid-state thermionic devices.

Equations 23 and 24 show that there are only 4 independent parameters determining the efficiency: $R_{tD}$, $R_D$, $\Phi_c$, and $\theta$. This is similar to the case of thermoelectric materials. Here $\Phi_c$ or more accurately $\Phi_c + 2$ is the average energy carried per carrier and the equivalent of a dimensionless Seebeck coefficient. Similar to thermoelectric materials, the parameters are related. In some cases, they cannot be tuned independently and in some other cases the optimum value of one parameter depends on the other parameters. For example as we increase the barrier height ($\Phi_c$), the average energy per carrier increases resulting in enhanced efficiencies. At the same time, as the barrier height increases, the Richardson current decreases, lowering the optimum power. Also at lower current values, larger thermal resistances are required to obtain reasonable efficiency values (as discussed before $J_c R_{tD}$ is the relevant parameter).

Figure 2 shows the efficiency divided by the Carnot efficiency for $\theta = 1.5$ (or $\eta_{carnot} = 33\%$) for three different electrical resistances and versus thermal resistance and barrier height. It also shows the optimum dimensionless power for the same electrical resistance values. As $R_D$ increases, the power drops as expected (note that the power is independent of the thermal resistance). $P_D$ (dimensionless power) of 0.01 is equal to power density of $p = 0.01 \frac{k_B T_a A_R T_a^2}{e}$. At room temperatures, this values is $p = 3\ kW.cm^{-2}$. We therefore can judge that the power values reported in figure 2 are in fact very large, 100 times more than reported values for vacuum-state thermionic generators. The reason is the substantially lower barrier heights. In terms of efficiency, as $R_D$ increases, the optimum value of $\Phi_c$ slightly shifts to larger values but overall it is somewhere between 3 and 5. So one can fix the value of $\Phi_c$ and try to optimize the values of resistances. Assuming the structure is fixed, and the contact layers (between cathode and barrier and between barrier and anode) are optimized to have the largest possible interfacial thermal resistance and lowest possible interfacial electrical resistance, the only parameter to adjust is the thickness of the barrier. We can assume that both electrical and thermal resistances increase linearly with increasing the thickness. Figure 3 shows that for $\Phi_c = 5$, the efficiency is almost independent of the choice of $R_D$ and therefore we can increase the thickness to increase the thermal resistance and the efficiency. This is expected as for this barrier height, the value of $J_c = 0.015$, which means that even for $R_D = 10$, the values of $J_c R_D$ is 0.15 which is considered to be much smaller than 0.5 and satisfies the criteria of $2 J_c R_D \ll 1$. The behavior is different for $\Phi_c = 4$. Here the efficiency slightly drops as $R_D$ increases (due to smaller $J_c$ values). As we increase the $R_{tD}$, $R_D$ also increases and therefore efficiency may or may not increase depending on the slope of $R_{tD}/R_D$.

Next, we show that it is also possible to find analytical solutions for optimum efficiency. The optimum current that results in the optimum efficiency can be estimated as:

$$J_D = J_c y_0 \tag{25}$$

$$\beta k = \frac{(\Phi_c \theta + 2)}{\left(2 + \frac{1}{J_c R_{tD}}\right)(\theta - 1)} \tag{26}$$

$$y_0 = 1 - \frac{1 + \beta k}{LW(x_1)} \tag{27}$$

$$x_1 = \theta^2 (1 + \beta k) \exp(J_c R_D (-2 + \beta k (-1 + \frac{1}{J_c R_D})) + \Phi_c (\theta - 1)) \tag{28}$$

Plugging in this current, we find efficiency to be:

$$\eta_{opt} = \frac{y_0 \ln\left(\frac{J_a}{J_0}\right) - J_c R_D y_0^2}{\left(2 + \frac{1}{J_c R_{tD}}\right)(\theta - 1) + y_0 (\Phi_c \theta + 2)} \tag{29}$$

We have plotted the results of equation 29 and we have not seen much difference between results of Eq. 29 and Eq. 24. Which means the results reported in figure 2 stay more or less the same if we try to optimize efficiency instead of power. Since the results are so similar, we do not plot them here. If we look back at figure 1, we see that this is expected since the optimum power and optimum efficiency occur at very close current values. The only exception to this is when the cathode work function value is close to the anode one. In that case, optimum efficiency and optimum power can occur at distinct currents. Houston[11] numerically showed that at optimum conditions, the following relation roughly holds between cathode and anode work function: $\Phi_c = \Phi_a$. For $\theta = 1.5$, which we used in our example, this means the work function of cathode is 1.5 times that of the anode. In this regime, our assumptions are correct.

Finally, we compare the performance the thermionic power generators with thermoelectric modules. If we set $\theta = 1.5$, or $\eta_{carnot} = 33\%$, the equivalent efficiency of an ideal thermoelectric modules, for ZT=1,2,3, $\eta_{TE}/\eta_{carnot}$ is equal to 0.19, 0.30 and 0.37 respectively. Figure 2 shows that using thermionic generators, and for the same $\theta$, efficiency values as high as $0.4\eta_{carnot}$ are achievable. In a large area of this graph, efficiency values are higher than 0.19. For example, if we set the cold side temperature to 300K and the hot side to 450K, we can recover $\theta = 1.5$. For this temperature range, there are only few thermoelectric materials available and bismuth antimony telluride is the best candidate.[12] Peak ZT value of p type bismuth antimony telluride is 1.8.[13] However, the average ZT of n-type and p-type bismuth antimony telluride in this temperature range is much smaller and close to 1, which means with the known thermoelectric modules, efficiency values higher than 20% of the Carnot efficiency are not achievable in this temperature range. But using thermionic modules (figure 2), such efficiency values are easily achievable. At the same time due to their smaller size, thermionic power generators own larger power density values and therefore they could be potentially a better alternative to thermoelectric modules.

Here we provide a discussion of the difference between our results and those obtained originally by Mahan. In his pioneering work,[2,14] Mahan assumed no internal resistance, small voltage and temperature differences over *symmetric* barriers, allowing him to use Taylor expansions and to address the problem within the linear transport regime. Our focus is the alternative regime where tall barriers and large temperature differences are of interest. We have also included the internal resistance which we believe could substantially lower the performance of the device. One question then is to what extent would our results tend to Mahan's. Since we do not assume a symmetric barrier and instead, we assumed the cathode work function to be larger than the anode work function, it is not trivial that we will recover Mahan's results even in the linear regime. However, upon plotting the results in this limit, we found that our results are very similar to Mahan's results. Figure 4 shows a comparison for small values of $\theta=1.05$ and zero internal resistance. The difference between our results and his increases as thermal resistance increases but as Fig. 4 shows the differences are not large. Interestingly, we observed that even when larger $\theta$ values are used, our results are close to Mahan's predictions. Therefore, as long as the internal electrical and thermal resistances are negligible, Mahan's formula can be used to estimate the performance of thermionic devices even at larger $\theta$ values. As we increase either $\theta$, $R_D$, or $R_t$, the optimum value of the barrier height tends to increase. Mahan's conclusion that the optimum barrier height is around $2\, k_B T$, therefore, is only valid in the linear regime. In the nonlinear regime and within reasonable values of thermal and electrical resistances, the optimum barrier height shifts to 3 to 5 $k_B T$. Finally, the reader should note that Mahan's conclusion that thermionic devices are not as good as thermoelectric devices, is only valid in the linear regime and if thermal resistance is dominated by the barrier lattice resistance. In the case that the thermal resistance of thermionic devices dominantly comes from the interfacial resistances, such conclusion cannot be drawn and thermionic convertors can be better than thermoelectric convertors.

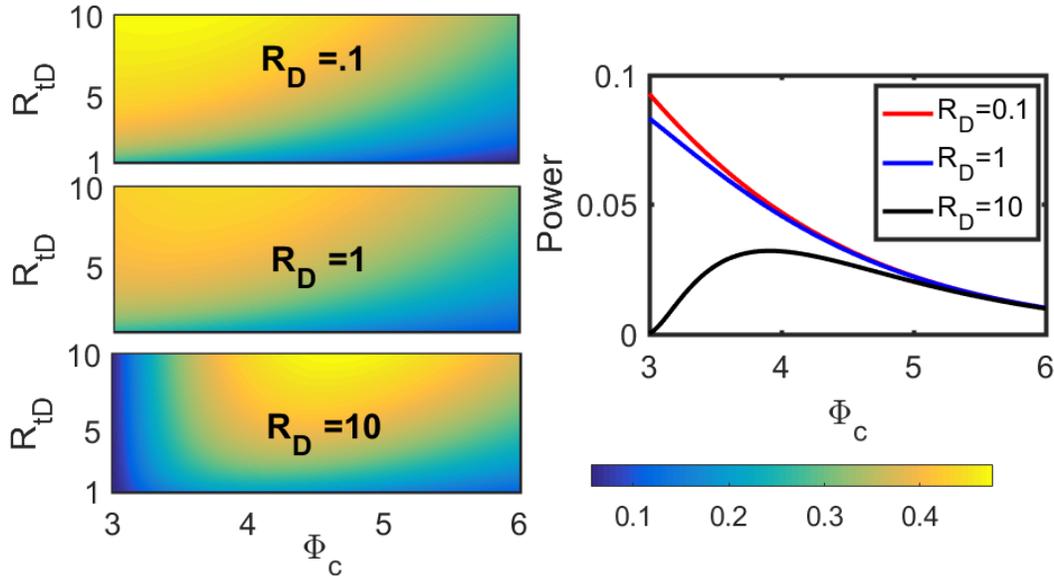

FIG. 2. Left: Efficiency at the optimum power calculated using Eq. 24 for $\theta = 1.5$ ($\eta_{Carnot} = 33\%$) and versus three main identified dimensionless parameters, $\Phi_c, R_{tD}\ and\ R_D$. Right: Optimum power (dimensionless) using Eq. 23, versus cathode's barrier height for three different $R_D$ values.

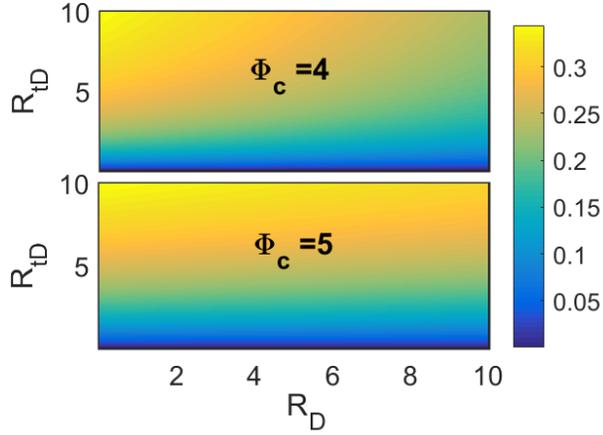

FIG. 3. Efficiency at the optimum power calculated using Eq. 24 for $\theta = 1.5$ and versus dimensionless thermal and electrical resistances. Here we fixed the barrier height close to the optimum value identified from figure 3, that is we fixed the barrier height to be 4 in the upper plot and 5 in the lower plot.

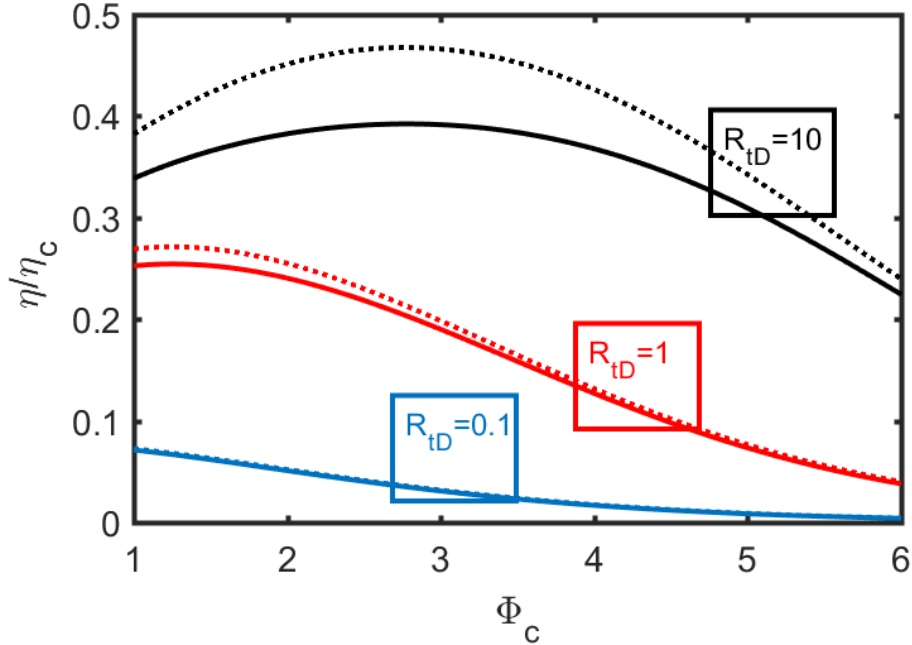

FIG. 4. Comparison of our results with those predicted by Mahan in the limit of zero internal resistance and small temperature differences: $\theta = 1.05$ and $R_D = 0$. Solid lines results from using Eq. 24 and dashed lines are plotted using equation 24 of ref. 1

### III. CONCLUSION

In summary, we have provided guidelines for design of solid state thermionic power generators. Although it is known that large thermal resistances and small electrical resistances are needed for these devices, here, we introduced quantitative measures to estimate the acceptable range of thermal and electrical resistances. If the internal electrical resistance is low, so that $J_c R_D < 0.5$, and if the internal thermal resistance is large so that $J_c R_{tD} > 1$, then the performance is not affected and there is no need to push the resistance values any further. We showed that similar to the case of thermoelectric modules, there are 4 independent parameters determining the efficiency: $\Phi_c, R_D, R_{tD}$, and $\theta$. We discussed that these parameters are related. In some cases, they cannot be independently tuned and in some other cases, their optimum value depends on the other parameters. We found the analytical expressions for the optimum load resistance as a function of the internal resistance which could be used when designing thermionic power generators for specific applications. When internal electrical is zero and for small values of thermal resistance, our results tend to those of Mahan, even under large temperature differences. However, we find that the optimum barrier height in the nonlinear regime is larger than $2 k_B T$ predicted by Mahan in the linear regime. This optimum shifts to larger values for larger resistances. Finally, we demonstrated that by choosing the right parameters, thermionic generators could achieve efficiencies higher than the state of the art thermoelectric modules.

## ACKNOWLEDGMENTS

The author would like to acknowledge Joseph A. King Jr (Arpa-e) for suggesting the problem and K. Esfarjani for his helpful feedback on the manuscript. This work is supported by the Air Force young investigator award, grant number FA9550-14-1-0316.

**Appendix A.** Lambert W function

Eq. 17 after neglecting the second order term can be written as:

$$\theta^2 \exp\bigl((\theta - 1)\Phi_c + 1 - 2J_c R_D\bigr) = \frac{J_c}{J_a} \exp\left(\frac{J_c}{J_a}\right) \qquad \text{A.1}$$

Solution of this equation is lambert W function.

$$\frac{J_c}{J_a} = LW\bigl(\theta^2 \exp\bigl((\theta - 1)\Phi_c + 1 - 2J_c R_D\bigr)\bigr) = LW(x_0) \qquad \text{A.2}$$

For small temperature differences $\theta \sim 1$ and negligible internal resistances, values of $x_0$ are close to e=2.7. Lambert W function of e is equal to 1 ($LW(e) = 1$) and it decreases to values smaller than 1 as we decrease $x_0$. This means $x_0$ values smaller than e are not of interest, since for such values leak current ($J_a$) is larger than the cathode current ($J_c$). For large temperature differences and tall barriers, values of $x_0$ can quickly increase. For example $\theta = 2, \Phi_c = 10$ and neglecting electrical resistance, gives $x_0 = 239497$. If needed, we can use simplified equations to estimate LW function. We found that using the following fits, one can accurately reproduce the LW function. However, the final equation for efficiency does not simplify enough to justify the use of these fits.

$$LW(x) = \begin{cases} \ln(x) - \ln(\ln(x)) + 0.3 & 10 < x \\ 0.58 \ln(x) + 0.4 & 2.7 < x < 10 \end{cases} \qquad \text{A.3}$$

## REFERENCES

(1) Mahan, G. D.; Sofo, J. O.; Bartkowiak, M. Multilayer Thermionic Refrigerator and Generator. *J. Appl. Phys.* **1998**, *83*, 17.

(2) Mahan, G. D.; Woods, L. M. Multilayer Thermionic Refrigeration. *Phys. Rev. Lett.* **1998**, *80*, 4016–4019.

(3) Shakouri, A.; LaBounty, C.; Abraham, P.; Piprek, J.; Bowers, J. E. Enhanced Thermionic Emission Cooling in High Barrier Superlattice Heterostructures. *MRS Proceedings*, 1998, *545*.

(4) Shakouri, A.; Bowers, J. E. Heterostructure Integrated Thermionic Coolers. *Appl. Phys. Lett.* **1997**, *71*, 1234.

(5) Wilson, V. C. Conversion of Heat to Electricity by Thermionic Emission. *J. Appl. Phys.* **1959**, *30*, 475–481.

(6) Hatsopoulos, G.; Kaye, J. Analysis and Experimental Esults of a Diode Configuration of a Novel Thermoelectron Engine. *Proc. IRE* **1958**, *46*, 1574–1579.

(7) Bifano, W. J.; Williams, R. M., J. Electrical Performance of a Rhenium-Niobium Cylindrical Thermionic Converter. *NASA-TN-D-4533* **1968**.

(8) Vining, C. B.; Mahan, G. D. The B Factor in Multilayer Thermionic Refrigeration. *J. Appl. Phys.* **1999**, *86*, 6852.

(9) MOSS, H. XXXII. Thermionic Diodes as Energy Converters†. *J. Electron. Control* **1957**, *2*, 305–322.

(10) Wang, X.; Zebarjadi, M.; Esfarjani, K. First Principles Calculations of Solid-State Thermionic Transport in Layered van Der Waals Heterostructures. *Nanoscale* **2016**, *8*, 14695–14704.

(11) Houston, J. M. Theoretical Efficiency of the Thermionic Energy Converter. *J. Appl. Phys.* **1959**, *30*, 481.

(12) Mehdizadeh Dehkordi, A.; Zebarjadi, M.; He, J.; Tritt, T. M. Thermoelectric Power Factor: Enhancement Mechanisms and Strategies for Higher Performance Thermoelectric Materials. *Mater. Sci. Eng. R Reports* **2015**, *97*, 1–22.

(13) Kim, S. Il; Lee, K. H.; Mun, H. A.; Kim, H. S.; Hwang, S. W.; Roh, J. W.; Yang, D. J.; Shin, W. H.; Li, X. S.; Lee, Y. H.; *et al.* Dense Dislocation Arrays Embedded in Grain Boundaries for High-Performance Bulk Thermoelectrics. *Science.* **2015**, *348*, 109–114.

(14) Mahan, G. D.; Sofo, J. O. The Best Thermoelectric. *Proc. Natl. Acad. Sci.* **1996**, *93*, 7436–7439.